\documentclass[conference,a4paper,10pt]{IEEEtran}
\ifCLASSOPTIONcompsoc
  \usepackage[nocompress]{cite}
\else
  \usepackage{cite}
\fi

\usepackage{graphicx}
\usepackage{multirow}
\usepackage{algorithm2e}
\usepackage{bm}
\usepackage{amssymb}
\usepackage{url}
\usepackage{caption}
\usepackage{subcaption}

\usepackage{color}

\newcommand{\Hide}[1]{}

\begin{document}
%
% paper title
% can use linebreaks \\ within to get better formatting as desired
\title{Performance and Security Evaluation of SDN Networks in OMNeT++/INET}

% author names and affiliations
% use a multiple column layout for up to two different
% affiliations
\author{\IEEEauthorblockN{Marco Tiloca}
\IEEEauthorblockA{SICS Swedish ICT AB, Security Lab\\
Isafjordsgatan 22, Kista (Sweden)\\
Email: marco@sics.se}
\and
\IEEEauthorblockN{Alexandra Stagkopoulou}
\IEEEauthorblockA{KTH Royal Institute of Technology\\
Isafjordsgatan 22, Kista (Sweden)\\
Email: stagk@kth.se}
\and
\IEEEauthorblockN{Gianluca Dini}
\IEEEauthorblockA{University of Pisa\\
Largo Lazzarino 1, Pisa (Italy)\\
Email: gianluca.dini@unipi.it}
}

% make the title area
\maketitle

\begin{abstract}
Software Defined Networking (SDN) has been recently introduced as a new communication paradigm in computer networks. By separating the control plane from the data plane and entrusting packet forwarding to straightforward switches, SDN makes it possible to deploy and run networks which are more flexible to manage and easier to configure. This paper describes a set of extensions for the INET framework, which allow researchers and network designers to simulate SDN architectures and evaluate their performance and security at design time. Together with performance evaluation and design optimization of SDN networks, our extensions enable the simulation of SDN-based anomaly detection and mitigation techniques, as well as the quantitative evaluation of cyber-physical attacks and their impact on the network and application. This work is an ongoing research activity, and we plan to propose it for an official contribution to the INET framework.
\end{abstract}


% For peer review papers, you can put extra information on the cover
% page as needed:
% \ifCLASSOPTIONpeerreview
% \begin{center} \bfseries EDICS Category: 3-BBND \end{center}
% \fi
%
% For peerreview papers, this IEEEtran command inserts a page break and
% creates the second title. It will be ignored for other modes.
\IEEEpeerreviewmaketitle

\section{Introduction}
\label{s:introduction}

In the recent years, \emph{Software Defined Networking} (SDN) \cite{SDNwhitepaper} has been more and more adopted as a new network communication paradigm \cite{Kreutz:2015}. Unlike the traditional Internet model, SDN separates the actual forwarding of network packets (data plane) from the management of network traffic and routes (control plane). In principle, a centralized \emph{SDN controller} determines how to handle different traffic segments, namely \emph{flows}, and installs related forwarding rules on simple \emph{switch} devices responsible for the actual packet forwarding. The SDN controller and the switches rely on a common set of APIs and control messages to interact with each other, so preventing interoperability issues among devices from different vendors. To this end, \emph{OpenFlow} \cite{OpenFlow} has become the de-facto protocol implemented in SDN controllers and switches. By entrusting all the monitoring and decision processes to the SDN controller, SDN considerably simplifies the management of large-scale networks. Also, it results in a faster and more flexible re-configuration of traffic patterns, if compared with traditional network architectures.

To deploy SDN networks that operate according to expectations, it is vital that, far before deployment, network designers can \emph{quantitatively} evaluate: i) network and communication performance; ii) effects and impact of security attacks against the network; and iii) accuracy and effectiveness of anomaly detection systems \cite{Giotis:2014}\Hide{\cite{Braga:2010}\cite{Mehdi:2011}}. To this end, network simulation represents a convenient and helpful tool to adopt, since it can be infeasible to perform the same evaluations in real, large-scale, networks. Especially for evaluating SDN-based monitoring systems and the impact of security attacks, it is convenient to perform such assessments at design time, so not interfering with the regular operations of real networks. On the other hand, an analytical approach is often infeasible, unless oversimplifying assumptions are made.

So far, there have been only a few contributions for simulation of SDN networks. Mininet is a common tool to perform functional testing of emulated networks based on OpenFlow \cite{Lantz:2010}. However, it focuses on real-time functional testing rather than on the simulation and evaluation of arbitrary network scenarios. The network simulator NS-3 provides an OpenFlow simulation model \cite{NS-3}. However, the SDN controller is not modeled as an external entity, and thus it is not possible to quantitatively evaluate the impact of the control channel or to consider multiple switches connected to the same SDN controller. More recently, \cite{Klein:2013} has proposed the implementation of OpenFlow components integrated in the INET framework \cite{Inet}, based on the network simulation environment OMNeT++ \cite{Omnet}. However, it models only the basic flow establishment between OpenFlow switches and a basic SDN controller. Also, it implements only flow-matching rules solely based on MAC address fields.

In this paper, we describe a set of extensions for the INET framework that enable performance and security evaluation in SDN networks. In particular, we have extended the model initially proposed in \cite{Klein:2013}, in order to support additional OpenFlow messages and enable the processing of network packets based on flow-matching rules of arbitrary complexity. Also, we have provided support for SDN-based monitoring systems, according to which the SDN controller: i) collects flow statistics from the connected switches; ii) analyzes collected samples in order to detect possible anomalies; and iii) possibly determines and disseminates policies to mitigate anomalies and restore normal operating conditions in the network. Finally, we have enabled the simulation of effects of security attacks in SDN network scenarios. To this end, we consider the INET-based attack simulation framework SEA++ that we previously described in \cite{SEA++paper} and whose functionalities have been adapted to support attack simulation in SDN networks. Intuitively, the user can describe different cyber-physical attacks by means of a high-level attack specification language, without altering the actual implementation of any of the INET software components. Events that reproduce the effects of the described attacks are injected at runtime during the simulation experiments. The approach devoted to reproducing and evaluating security attacks is not strictly related to SDN, i.e. it can in principle be reused together with any network architecture and scenario sopported by the INET framework.

We believe that our extended simulation tool allows researchers and network designers to effectively and conveniently evaluate SDN architectures at design time, both in attack-free scenarios and in case different security attacks are performed. In particular, it makes it possible to evaluate an SDN scenario in terms of: i) network and communication performance in an attack-free case; ii) effectiveness and reactiveness of SDN-based monitoring systems; iii) quantitative effects of security attacks, as to how attacks affect performance indicators in the same network scenario; and iv) quantitative effectiveness of security countermeasures. This work is an ongoing research activity, and we plan to propose it for an official contribution to the INET framework. Our extended simulation framework is currently under development, and the source code is available at \cite{SDNsimulator}. 

The paper is organized as follows. Section \ref{s:sdn} overviews Software Defined Networking. Section \ref{s:extensions} describes our extensions to the INET framework which support SDN, OpenFlow, SDN-based monitoring systems, and simulation of effects of security attacks. In Section \ref{s:results}, we evaluate a simple Denial of Service attack against a server host, and present preliminary results. Finally, Section \ref{s:conclusion} concludes the paper and anticipates future works and research directions.

\section{Packet forwarding in SDN}
\label{s:sdn}

SDN essentially relies on the separation of \emph{data plane} and \emph{control plane}. In particular, the data plane is entrusted to simple \emph{switches} that forward network packets according to stored \emph{flows} and related matching rules. The control plane is entrusted to a centralized SDN \emph{controller}, which establishes packet flows and installs them on the switches. The SDN controller and switches interact with each other through dedicated control messages and APIs, such as the ones provided by the common \emph{OpenFlow} protocol \cite{OpenFlow}.
\begin{figure}[tbp]
\centering
\includegraphics[width=.95\linewidth]{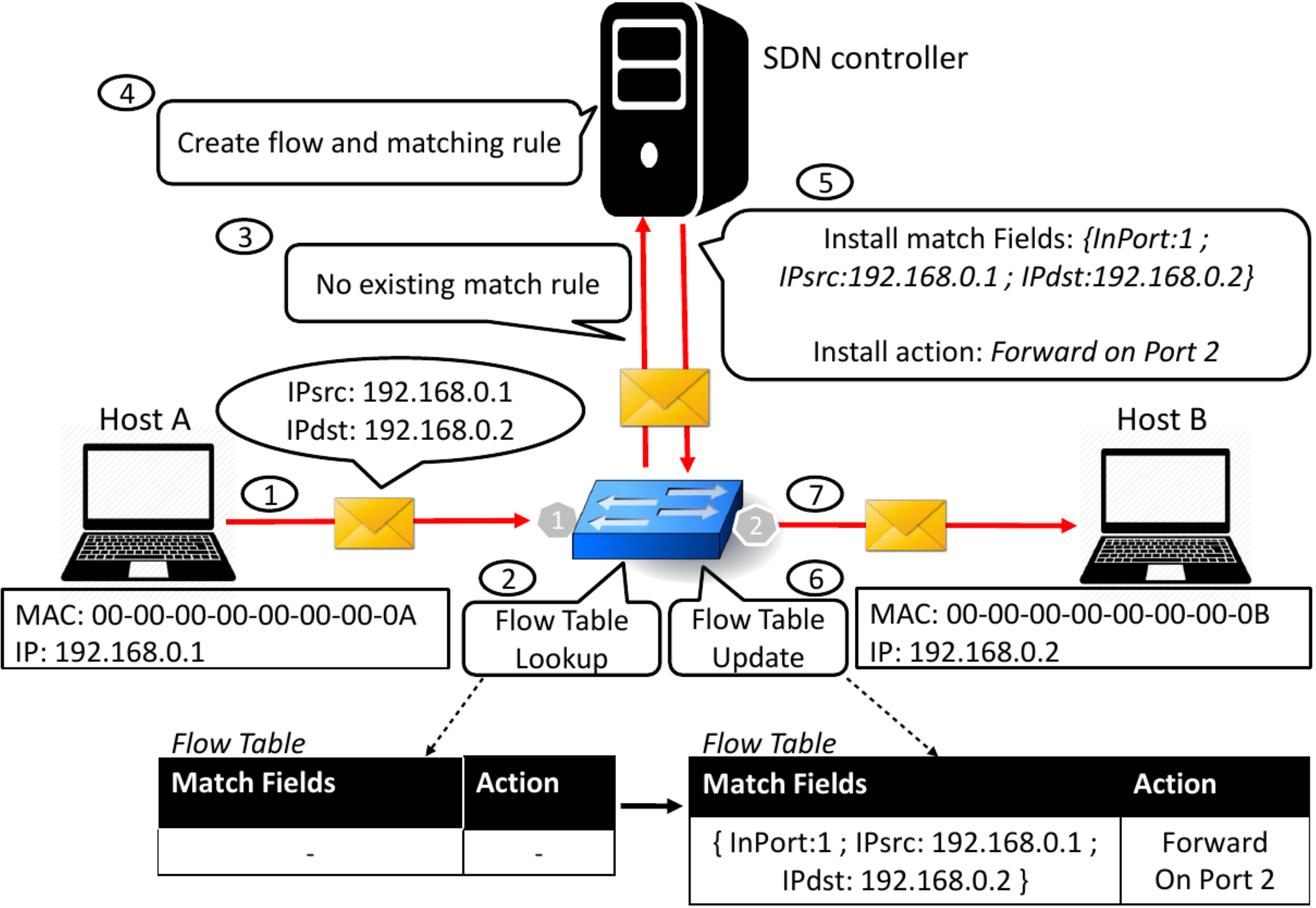}
\caption{Flow establishment and packet forwarding.}
\label{fig:scenario}
\end{figure}

Figure \ref{fig:scenario} shows an example of flow establishment and packet delivery. First, host A sends a packet P to host B (step 1). Upon receiving packet P, the switch looks for a possible match between P and the flows installed in its flow table (step 2). If no matching rule to process packet P is found, the switch asks the SDN controller for further instructions (step 3). Then, the SDN controller creates a new flow (step 4), and installs the related matching rule and actions on the switch (step 5). That is, it instructs the switch that all packets coming from Port 1, with IP source address 192.168.0.1 and IP destination address 192.168.0.2 must be sent out over Port 2. After that, the switch installs the new flow on its flow table, as a pair of matching fields and actions to be performend on any packet matching with that flow (step 6). If more output ports are available, the switch is initially instructed to send out this first packet P over all ports different than Port 1. Later on, upon receiving a reply packet on a specific port, the switch contacts again the SDN controller, which modifies the action associated to that flow by specifying the exact outgoing port to consider. Finally, the switch forwards packet P to host B (step 7).
\section{SDN extensions for INET}
\label{s:extensions}

This section overviews our extensions for the INET framework. Section \ref{ss:SDNsupport} presents the support for flow establishment and packet forwarding based on OpenFlow. Section \ref{ss:SDNmonitoring} presents the support for SDN-based monitoring systems. Section \ref{ss:attacks} presents the support for the evaluation of security attacks. Our extended simulation framework is under development, and the source code is available at \cite{SDNsimulator}.
\subsection{Support to SDN architecture and OpenFlow}
\label{ss:SDNsupport}

SDN relies on two fundamental elements: i) the SDN controller and switches; and ii) the exchange of OpenFlow messages to establish flows and install them on the switches.

We have considered the model initially proposed in \cite{Klein:2013}, that provides a number of essential OpenFlow messages and the implementation of the switches and SDN controller nodes. In particular, the SDN controller is essentially a host running an application which relies on a typical TCP/IP stack and models a specific controller behavior. Instead, the switches are modelled as a new type of node, where a control plane running a TCP application on top of a TCP/IP stack interacts with multiple data plane instances, by means of the OMNeT++ signal concept. Both the SDN controller and the switches rely on a specific time model to take into account the processing time of real OpenFlow units.
\begin{figure}[tbp]
\centering
\includegraphics[width=.95\linewidth]{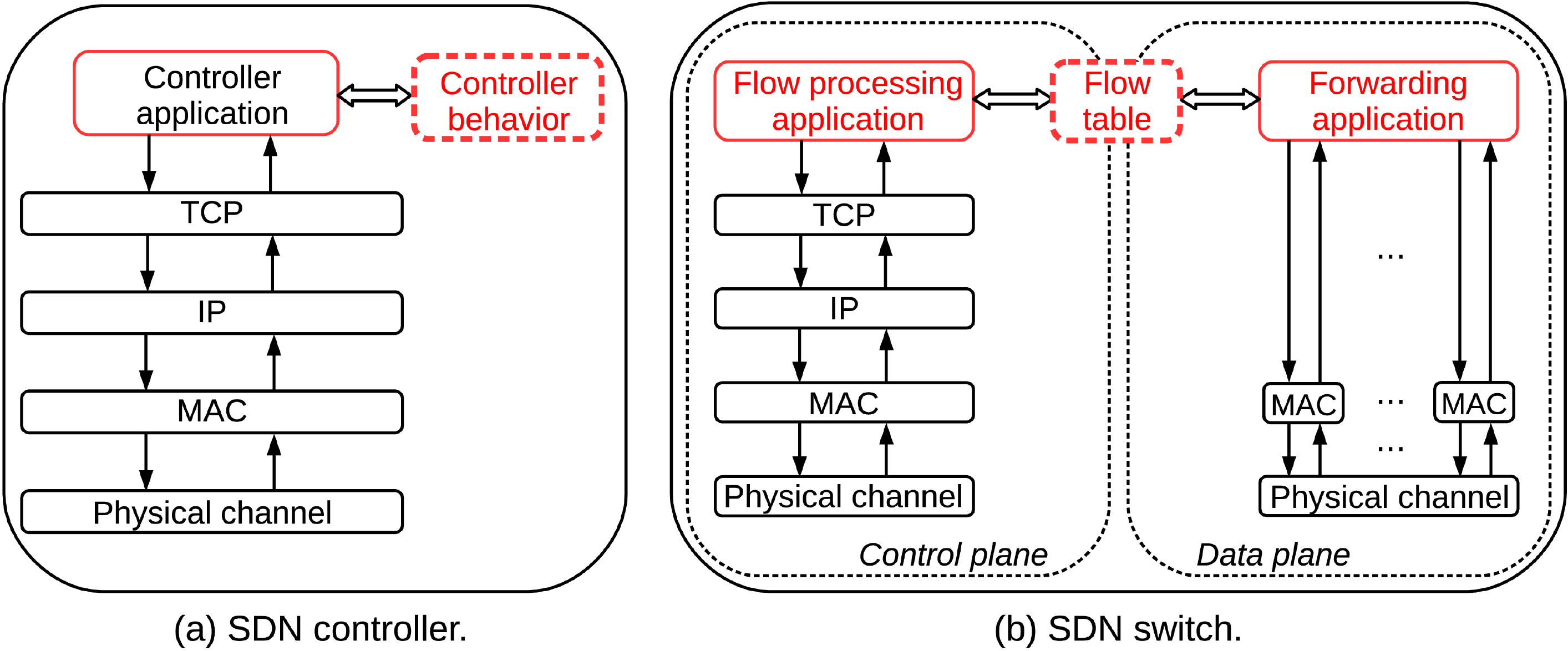}
\caption{Overview of a SDN controller and switch.}
\label{fig:switch_and_controller}
\end{figure}

We extended the model implemented in \cite{Klein:2013}, providing additional functionalities that enable the evaluation of SDN-based monitoring systems and impact of security attacks (see Sections \ref{ss:SDNmonitoring} and \ref{ss:attacks}). In particular, we provided additional support to: i) exchange and process the OpenFlow control messages OFPT\_STATS\_REQUEST and OFPT\_STATS\_REPLY between the SDN controller and the switches for collection of flow statistics; ii) exchange and process the OpenFlow messages OFPT\_FLOW\_REMOVED sent by the switches to the SDN controller to report expired flows; and iii) allow the switches to perform the matching of incoming packets with installed flows based on arbitrary packet fields (rather than on MAC addresses only).
\Hide{
\begin{figure}
\centering
\begin{subfigure}{.5\columnwidth}
  \centering
  \includegraphics[width=.99\linewidth]{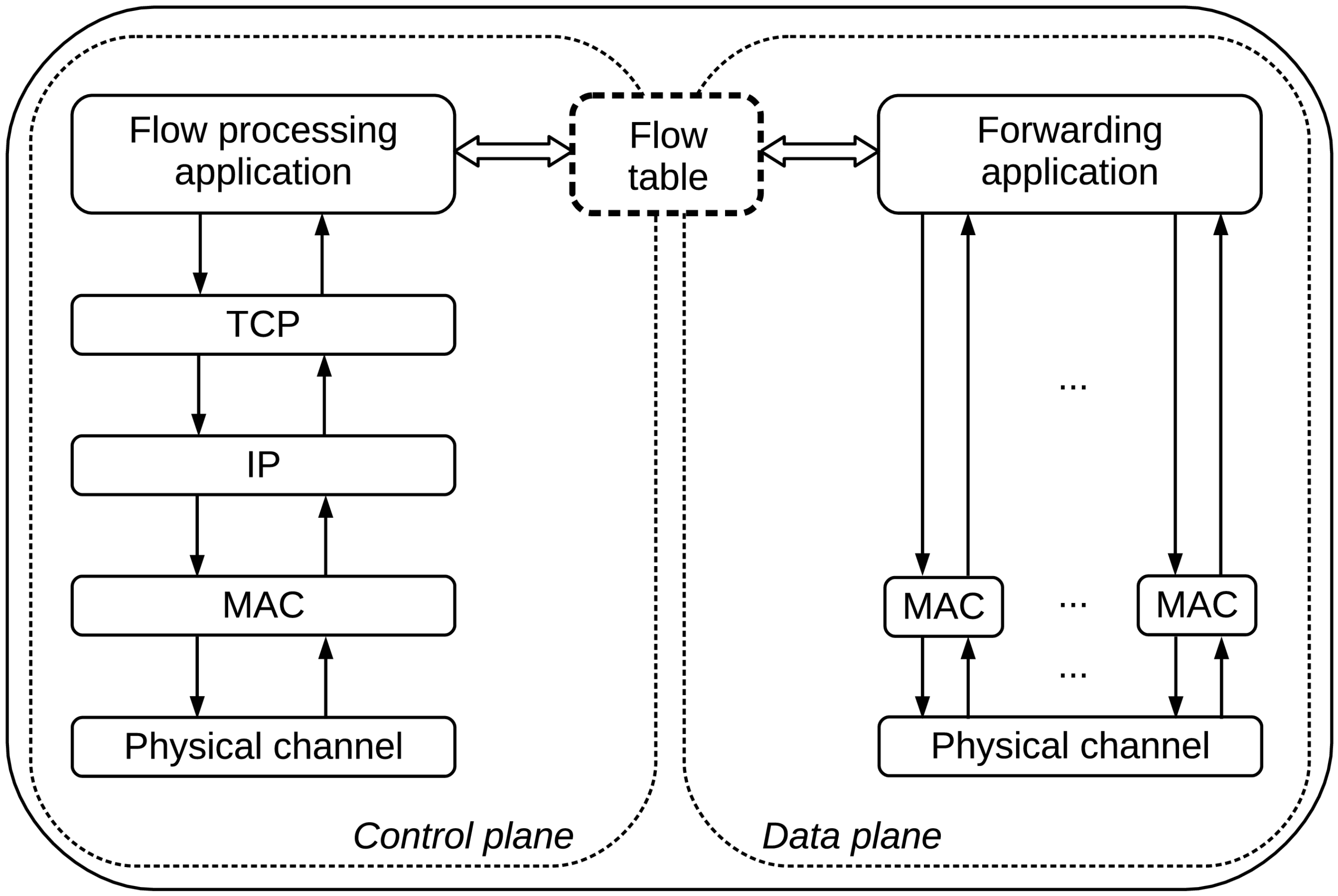}
  \caption{OpenFlow switch.}
  \label{fig:switch}
\end{subfigure}%
\begin{subfigure}{.5\columnwidth}
  \centering
  \includegraphics[width=.48\columnwidth]{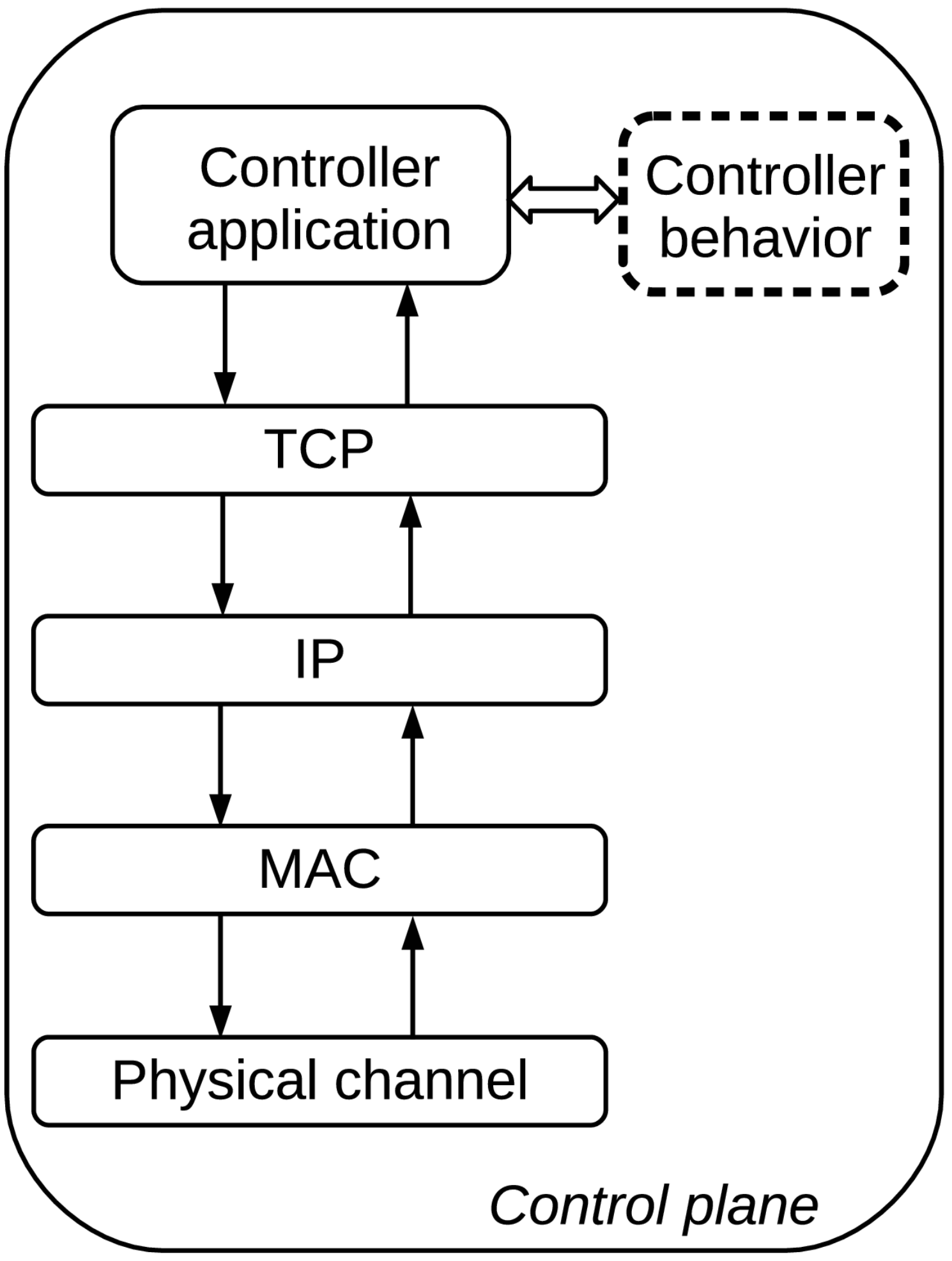}
  \caption{OpenFlow controller.}
  \label{fig:controller}
\end{subfigure}
\caption{Overview of an OpenFlow switch and controller.}
\label{fig:switch_and_controller}
\end{figure}
}

Figure \ref{fig:switch_and_controller} shows the architectural overview of the SDN controller (a) and a SDN switch (b). The elements coloured in red have been extended in our implementation, in order to enable statistic collection and network monitoring, as well as the reporting of flow expiration and the arbitrary-complex matching of packets with flows installed on switches. The SDN controller is simply modelled as a generic host, running a traditional TCP/IP stack. Then, a specific \emph{Controller application} is responsible for the establishment of flows, and their installation, update and revocation on the switches. Of course, the \emph{Controller application} can be entrusted with additional services, such as network monitoring and anomaly detection (see Section \ref{ss:SDNmonitoring}). Policies, algorithms and parameters according to which the \emph{Controller application} behaves are specified in the \emph{Controller behavior} module.

The switch is composed of two different segments, sharing the same \emph{Flow table}. That is, the \emph{Control plane} is also modelled as a typical TPC/IP stack, through which a \emph{Flow processing application} can exchange control messages with the SDN controller. Instead, the \emph{Data plane} is a set of minimal communication stacks, each one relying on a dedicated MAC interface. Then, all MAC interfaces are connected to the same \emph{Forwarding application}, which forwards packets from an incoming MAC interface to an outgoing MAC interface, according to the flow-matching rules and related forwarding actions stored in the \emph{Flow table}. If no matching is produced, the \emph{Forwarding application} asks the \emph{Flow processing application} to contact the SDN controller and establish a new flow, before proceeding.

The following OpenFlow messages are considered in order to support the establishment and updates of flows.\\
\noindent
$\bullet$ OFPT\_PACKET\_IN. Sent by the switch to the SDN controller, when a packet is received and no match is produced.\\
\noindent
$\bullet$ OFPT\_PACKET\_OUT. Sent by the SDN controller to a switch, specifying to send a packet over a specific interface.\\
\noindent
$\bullet$ OFPT\_FLOW\_MOD. Sent by the SDN controller to a switch, specifying to install/modify a flow in its flow table.\\
\noindent
$\bullet$ OFPT\_FLOW\_REMOVED. Sent by a switch to the SDN controller, notifying that a flow in the flow table has expired.
\subsection{SDN-based monitoring systems}
\label{ss:SDNmonitoring}

The SDN controller can run additional application services to perform security monitoring of the network. This practically relies on three modules, namely \emph{flow statistic collection}, \emph{anomaly detection}, and \emph{anomaly mitigation}. That is, the SDN controller periodically sends an OFPT\_STATS\_REQUEST OpenFlow message to the switches, according to a pre-configured polling interval. The switches reply with an OFPT\_STATS\_REPLY OpenFlow message, reporting the packet matches and the accesses to their flow tables occurred during the current time window. Given this information, the SDN controller can analyze the collected statistics, and look for possible anomalies or ongoing attacks, such as Denial of Service (DoS) or wormhole propagation. The actual anomaly detection process can rely on several different techniques, e.g. machine learning \cite{Ahmed:2007}, data mining \cite{Wu:2009}, or entropy based algorithms \Hide{\cite{Lakhina:2005}}\cite{Mousavi:2015}\cite{Oshima:2010}.

When the SDN controller identifies traffic anomalies or ongoing attacks, it performs mitigating actions to limit or neutralize their impact. That is, the SDN controller sends OFPT\_FLOW\_MOD messages to specific switches, to install or update flows in their flow tables. Such flows and related policies aim at blocking malicious traffic, e.g. by dropping or caching packets that are addressed to presumed victim hosts or coming from suspected attack sources.
\subsection{Simulation of security attacks}
\label{ss:attacks}
Our extensions allow network designers to \emph{quantitatively} evaluate the impact and effects of security attacks against SDN networks, i.e. how attacks affect performance indicators with respect to the same scenario in the attack-free case. This makes it possible to rank different attacks according to their severity, and hence to easier select effective countermeasures to adopt. Rather than executing security attacks by implementing their actual performance, we \emph{reproduce} their effects against the network and applications. Evaluation of security attacks relies on two fundamental components, i.e. a high-level \emph{Attack Specification Language} and an \emph{Attack Simulation Engine}, described in Sections \ref{sss:asl}-\ref{sss:injection}.

We previously presented an earlier version of such components as part of the INET-based attack simulation framework SEA++ \cite{SEA++paper}, and adapted their functionalities to support attack simulation in SDN networks. Note that the approach adopted to reproduce and evaluate security attacks is not strictly related to SDN, i.e. it can in principle be reused for any network architecture sopported by INET. At the same time, our extended simulation tool makes it possible to evaluate the impact of security attacks that specifically consider switches as actual attack victims (e.g. injection of fake flows to install) or compromised units contributing to the attack execution (e.g. through packet dropping or replication). Our implementation activity is currently focused on enabling the evaluation of such attacks involving switches\Hide{ and, possibly, the SDN controller}.

\subsubsection{Attack description}
\label{sss:asl}
The Attack Specification Language (ASL) allows the user to describe attacks to be evaluated, in terms of their final \emph{effects}. That is, the user assumes that attacks can be successfully performed, regardless how an adversary can specifically mount and execute them. Then, the user describes attacks as sequence of events that atomically take place during the network simulation. To this end, the ASL provides a collection of \emph{primitives} organized into two sets, i.e. \emph{node primitives} and \emph{message primitives}.

Node primitives account for \emph{physical attacks} against network nodes\Hide{ and are used to describe alterations in nodes' behavior}. A physical attack is composed by a single node primitive. The following node primitives are available:\\

\noindent
$\bullet$ \textsf{destroy(nodeID, t)} - Remove node '\textsf{nodeID}' from the network at time '\textsf{t}', after which it cannot take part to network communication any longer.\\
\noindent
$\bullet$ \textsf{move(nodeID, t, x, y, z)} - Change the current position of node '\textsf{nodeID}' to a new position \{\textsf{x,y,z}\} at time '\textsf{t}'.\\

Message primitives account for \emph{cyber attacks} and describe actions on network packets. Packet fields are addressed by means of the \emph{dot} notation \textsf{packet.layer.field}. \Hide{Thus, the user has to know the specific protocols composing the adopted communication stack, including their packet header's structure. }The following node primitives are available:\\

\noindent
$\bullet$ \textsf{drop(pkt)} - Discards the packet '\textsf{pkt}'.\\
\noindent
$\bullet$ \textsf{create(pkt, fld, content, ...)} - Creates a new packet '\textsf{pkt}' and fill its field '\textsf{fld}' with '\textsf{content}'. It is possible to specify the content of multiple fields through a single invocation.\\
\noindent
$\bullet$ \textsf{clone(srcPkt, dstPkt)} - Produces a perfect copy '\textsf{dstPkt}' of the packet '\textsf{srcPkt}'.\\
\noindent
$\bullet$ \textsf{change(pkt, fld, newContent)} - Writes '\textsf{newContent}' into the field '\textsf{fld}' of packet '\textsf{pkt}'.\\
\noindent
$\bullet$ \textsf{send(pkt, d)} - Schedules the transmission of a packet '\textsf{pkt}' produced by '\textsf{clone()}' or '\textsf{create()}', after a delay '\textsf{d}'.\\
\noindent
$\bullet$ \textsf{retrieve(pkt, fld, var)} - Assigns the content of the field '\textsf{fld}' of packet '\textsf{pkt}' to the variable '\textsf{var}'.\\
\noindent
$\bullet$ \textsf{put(pkt, dstNodes, TX \textbar\ RX, updateStats, d)} - Inserts the packet '\textsf{pkt}' either in the TX or RX buffer of all nodes in the '\textsf{dstNodes}' list, after a delay '\textsf{d}'.\\

The ASL provides statements to specify \emph{conditional attacks}, i.e. lists of events described through message primitives that occur on a declared list of nodes if a condition is evaluated as TRUE. That is, as a general example:
\begin{verbatim}
from T nodes = <list of nodes> do {
  filter(<condition>) <list of events>
}
\end{verbatim}

Also, it is possible to specify \emph{unconditional attacks}, i.e. list of attack events described through message primitives, and reproduced on a periodical fashion or upon the occurrence of specific conditions evaluated by network nodes at runtime. For instance, the statement \verb+from T every P do {<list of events>}+ specifies that the list of events takes place periodically on the declared list of nodes, since time T and with period P.

\subsubsection{Attack Simulation Engine}
\label{sss:engine}
After having described the attacks to be evaluated, the user simply runs a simulation campaign on the enhanced INET framework, in order to evaluate the impact and effects of the described attacks. To this end, the \emph{Attack Simulation Engine} (ASE) considers network nodes as implemented by an \emph{Enchanced Network Node} module. The latter is in turn composed of: i) an \emph{Application} module possibly including different sub-modules modelling the actual node application(s); ii) an arbitrarily complex collection of protocols composing the communication stack; and, finally iii) a \emph{Local Event Processor} (LEP) module. Notice that all such modules but LEP can be off-the-shelf.

The LEP module manages the attack events and operates transparently with respect to the other components of the \emph{Enhanced Network Node} module. Specifically, the LEP module intercepts incoming and outgoing network packets traveling through a node's communication stack, acting as \emph{gate-bypass} between each pair of INET modules implementing the different communication layers. Then, depending on the considered attacks to be evaluated, it can inspect and alter packets' content, inject new packets, or even discard intercepted ones. Finally, the LEP module can also alter the node's behavior at different layers, change its position in space, or even neutralize the node by making it inactive.

To address the presence of multiple network nodes and enable the simulation of complex attacks, we instantiate an \emph{Enhanced Network Node} module for each network node, and a single \emph{Global Event Processor} (GEP) module that connects all the \emph{Enhanced Network Node} modules with one another. The GEP module is separately connected with every LEP module, so allowing them to synchronize and communicate with one another in order to implement more complex, possibly distributed, security attacks. Finally, the LEP and GEP modules handle packets at different communication layers and conveniently access their header fields by means of the OMNeT++ \emph{descriptor} classes.
\begin{figure}[tbp]
\centering
\includegraphics[width=2.7in]{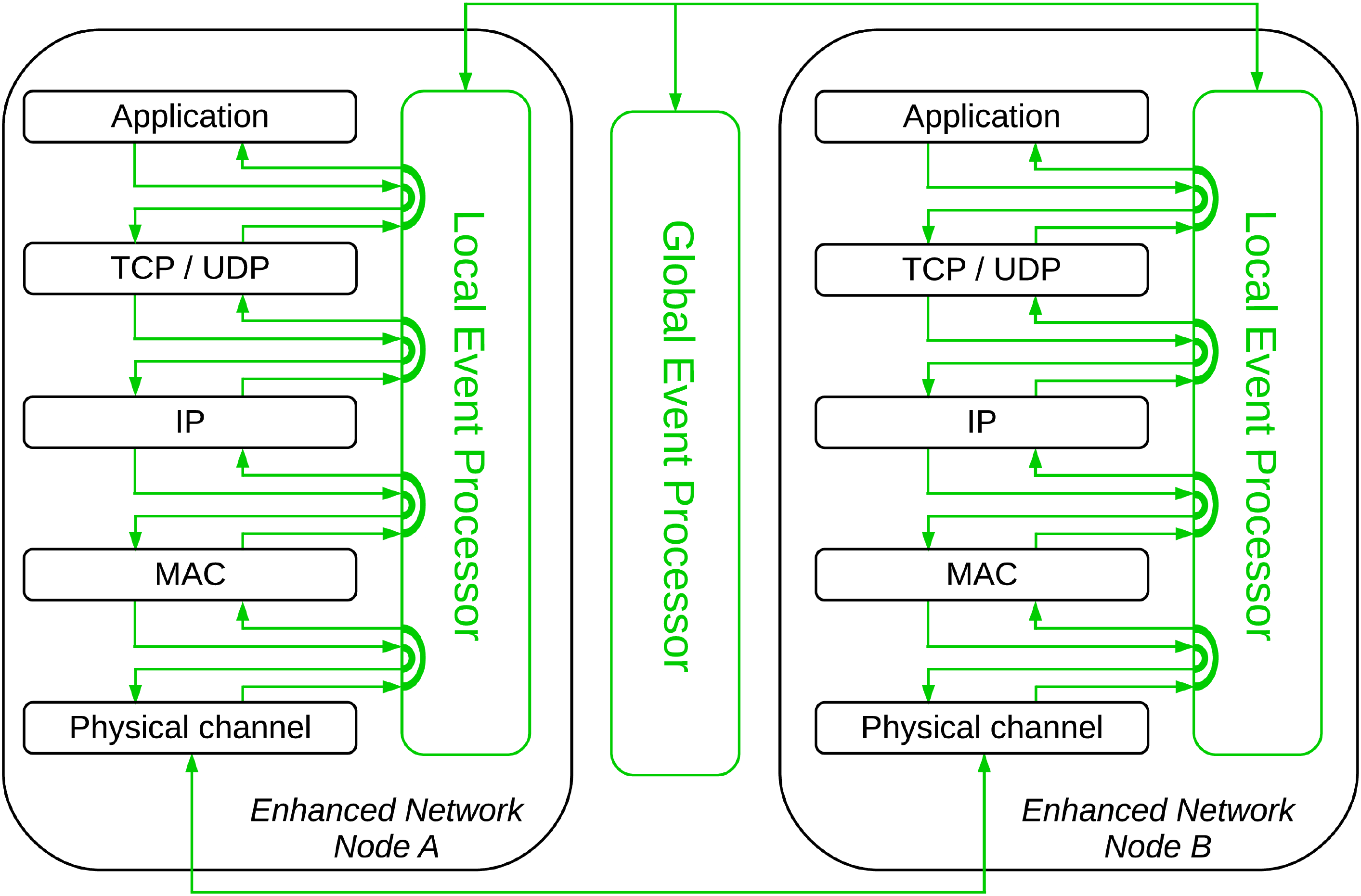}
\caption{Architecture of the \emph{Attack Simulation Engine}.}
\label{fig:attack_simulator}
\end{figure}

Figure \ref{fig:attack_simulator} shows the overall architecture of the ASE, with reference to two interconnected network nodes. Our extensions integrate the \emph{Local Event Processor} and \emph{Global Event Processor} modules highlighted in green within the INET framework, in order to correctly manage simulation events and network packets. This requires particular attention for the SDN switches, to maintain the separation between the \emph{Control plane} and the \emph{Data plane}, and to correctly manage the multiple MAC interfaces (see Figure \ref{fig:switch_and_controller}(b)).

Note that the ASE consists in additional components integrated within the INET framework to support the processing of attack events. That is, we do \emph{not} fundamentally modify INET as to the handling and scheduling of simulation events, and we do \emph{not} modify any of the available applications, communication protocols, or physical models. Most important, the user is \emph{not} required to implement or customize any component of the simulation platform.

\subsubsection{Injection of attack events}
\label{sss:injection}
The attack description based on ASL is converted into a XML configuration file by means of a Python \emph{Attack Specification Interpreter}, and then provided as input to INET upon simulation startup. Such configuration file is composed of three different sections, i.e a first part listing all the specified physical attacks, a second part listing all the specified conditional attacks, and a final third part listing all the specified unconditional attacks. At simulation startup, the ASE parses the XML configuration file and proceeds as follows. For each node $n$ involved in at least one attack, the ASE:\\
\noindent
$\bullet$ Creates one list $LP_n$, each element of which includes the description of one physical attack involving node $n$. The list elements are cronologically ordered according to the respective attack's occurrence time.\\
\noindent
$\bullet$ Creates one list $LC_n$, each element of which includes the description of one conditional attack involving node $n$. The list elements are cronologically ordered according to the respective attack's starting time.\\
\noindent
$\bullet$ Creates one list $LU_n$, each element of which includes the description of one unconditional attack involving node $n$. The list elements are cronologically ordered according to the respective attack's starting time.\\

After that, the ASE starts a number of timers, each one associated to a specified attack. That is, for each node $n$ involved in at least one attack, the ASE:\\
\noindent
$\bullet$ Creates a set of attack timers $TP_n$, each one of which associated to one physical attack involving node $n$.\\
\noindent
$\bullet$ Creates a set of attack timers $TC_n$, each one of which associated to one conditional attack involving node $n$.\\
\noindent
$\bullet$ Creates a set of attack timers $TU_n$, each one of which associated to one unconditional attack involving node $n$.\\
\noindent
$\bullet$ Starts all the timers in $TP_n$, $TC_n$, and $TU_n$, in order to schedule the respective attack's occurrence.\\

Throughout the network simulation, the ASE proceeds as follows. When an attack timer associated to a node $n$ expires, the ASE retrieves the associated attack $A$. Then:\\
\noindent
$\bullet$ If $A$ is a physical attack, the ASE executes the associated node primitive, and removes $A$ from the attack list $LP_n$.\\
\noindent
$\bullet$ If $A$ is a conditional attack, from then on the ASE starts intercepting packets flowing through node $n$'s communication stack, by means of node $n$'s LPE. Intercepted packets are filtered, based on the condition specified in the conditional statement of attack $A$. For each packet that satisfies the conditional statement, the ASE executes the list of events described by the message primitives in $A$. The execution of some node primitives may involve also the GEP as well as the LEP modules of other nodes than $n$.\\
\noindent
$\bullet$ If $A$ is an unconditional attack, from then on the ASE starts executing the corresponding list of message primitives, and repeatedly performs $A$ according to the occurrence frequency in the attack description. The GEP is responsible for starting the actual reproduction of uncondintional attacks.
\section{Evaluation of a Denial of Service attack}
\label{s:results}
\begin{figure}[bp]
\centering
\includegraphics[width=2.6in]{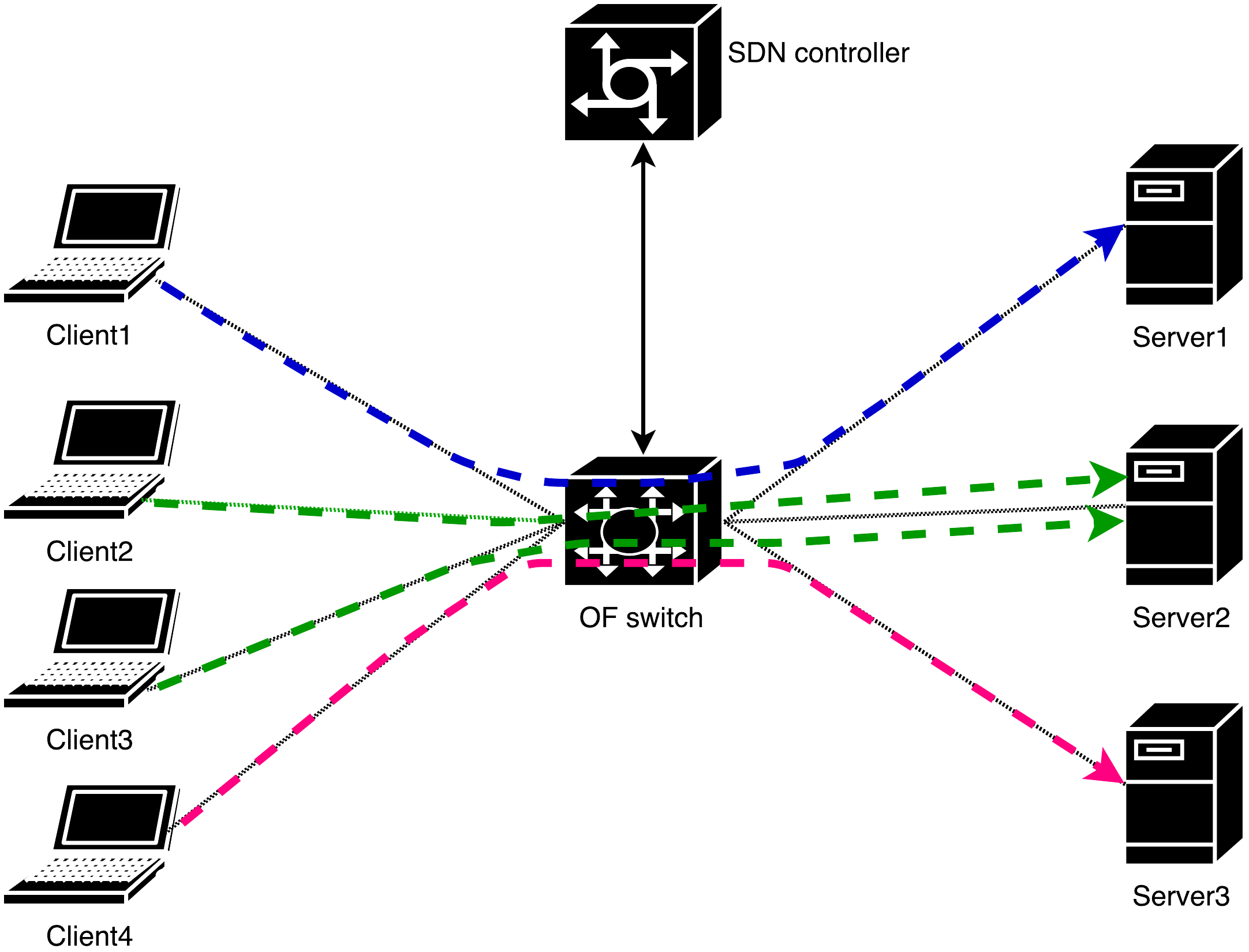}
\caption{Evaluated SDN network scenario.}
\label{fig:experiment_scenario}
\end{figure}

In this section, we simulate a simple Denial of Service attack against a server host, and present some preliminary results. We refer to the scenario in Figure \ref{fig:experiment_scenario}, which includes: i) a SDN controller and a switch; and ii) four client hosts and three server hosts, running a UDP application. Besides, Client1 sends $10$ packets per second to Server1; Client2 and Client3 send $5$ and $3.33$ packets per second to Server2, respectively; Client4 sends $5$ packets per second to Server3.

Each flow installed on the switch expires every $30$ seconds. When this happens, the switch notifies the SDN controller, removes the flow and possibly re-establishes it upon receiving new packets from/to the involved host(s). The SDN controller periodically collects flow statistics from the switch, according to a configurable interval $I$. Besides, the SDN controller runs a monitoring application that analyzes flow statistics in order to detect possible traffic anomalies. In particular, it relies on: i) entropy-based techniques for anomaly detection \cite{Mousavi:2015}\cite{Oshima:2010}; and ii) bounded rates for transmission/reception of packets on a single-node basis. If traffic anomalies are detected on a given flow, the SDN controller sends an OpenFlow message OFPT\_FLOW\_MOD to the switch, in order to install a selective \emph{drop} policy and discard all packets matching with that flow until further notice. In this scenario, we consider an adversary that has compromised Client3, and exploits it to transmit \emph{additional} network packets to Server2, starting from time $t = 90$ s, and according to a packet injection rate $R$.

\begin{figure}[tbp]
\centering
\includegraphics[width=2.7in]{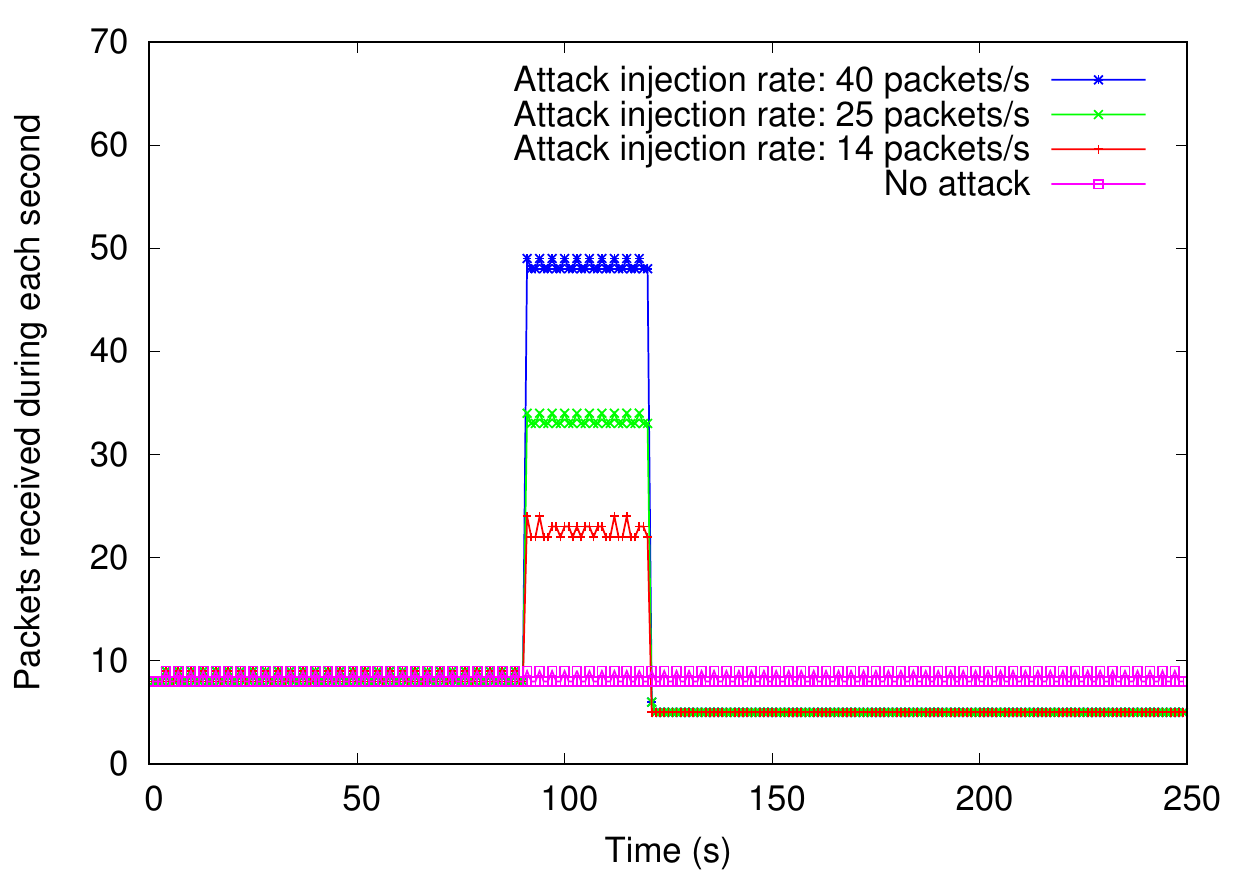}
\caption{Packet reception on Server2 ($I$=$30$ s).}
\label{fig:fixed_polling_interval_30s}
\end{figure}

\begin{figure}[tbp]
\centering
\includegraphics[width=2.7in]{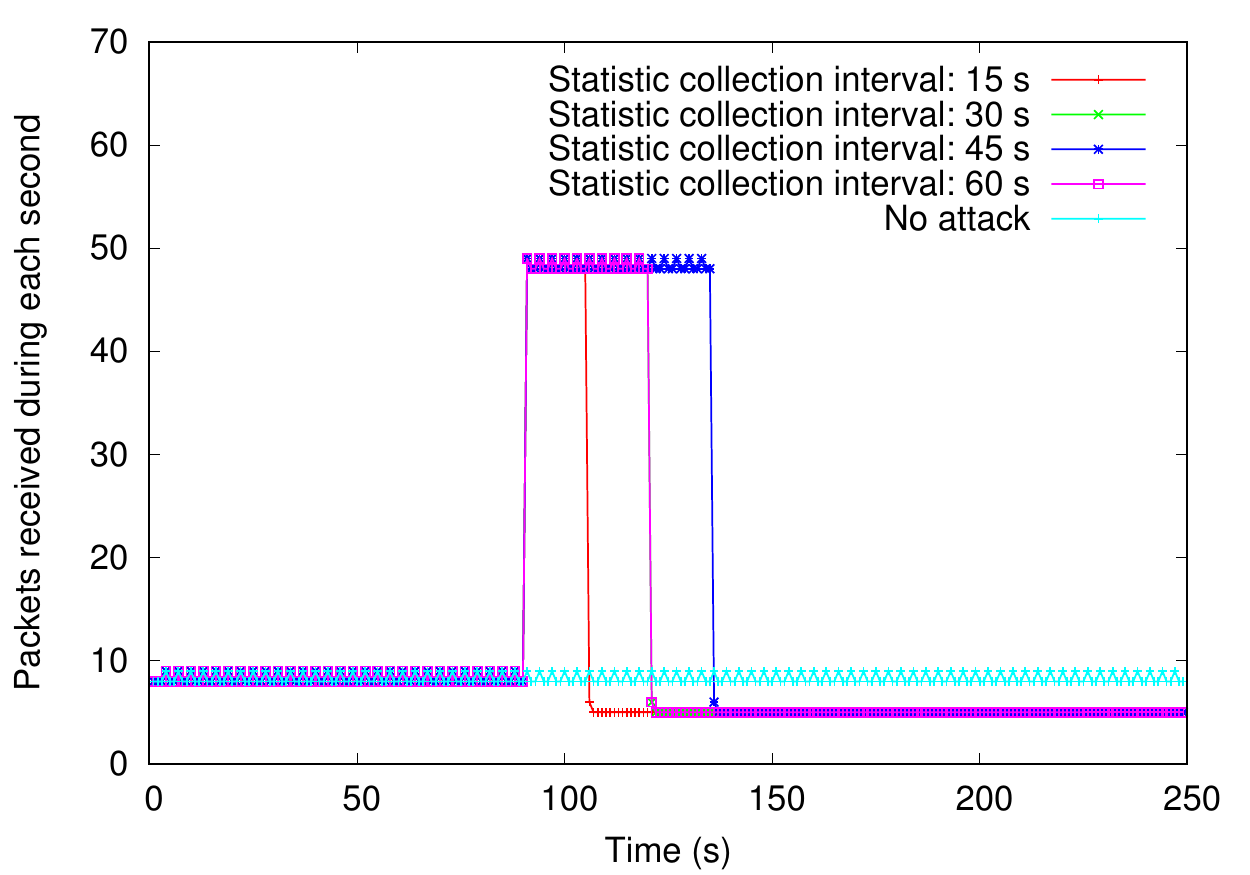}
\caption{Packet reception on Server2 ($R$=$40$ pkts/s).}
\label{fig:fixed_attack_rate_40pkt_s}
\end{figure}

Figures \ref{fig:fixed_polling_interval_30s} and \ref{fig:fixed_attack_rate_40pkt_s} show the packet reception on Server2, considering the attack-free case ``No attack'' as baseline. A value plotted at second $t = s$ indicates the number of packets overall received by Server2 during the last second, i.e. between $t = s-1$ and $t = s$. Specifically, Figure \ref{fig:fixed_polling_interval_30s} considers different injection rates $R$, and shows that the greater $R$ the more (attack) packets are received by Server2, until the traffic anomaly is mitigated at $t = 120$ s. From then on, the switch discards all packets coming from Client3 and addressed to Server2. Figure \ref{fig:fixed_attack_rate_40pkt_s} considers different statistic collection intervals $I$, and shows the different times that it takes to detect and mitigate the attack, depending on the interval $I$ considered by the SDN controller.
\section{Conclusion}
\label{s:conclusion}

We have presented our extensions to the INET framework to support the evaluation of performance and security attacks in SDN scenarios. Our extensions allow the user to evaluate performance of a SDN architecture, assess accuracy and reactiveness of SDN-based monitoring systems, and quantitatively evaluate the impact of security attacks. We have evaluated a simple Denial of Service attack, and presented preliminary results. This work is an ongoing research activity, and we plan to propose it for an official contribution to the INET framework. Future work will focus on evaluating different classes of security attacks, considering different SDN-based monitoring systems and adversary models.
\section*{Acknowledgments}
\label{s:acknowledgment}
The authors would like to sincerely thank the anonymous reviewers and the shepherd Michael Kirsche for their insightful comments and suggestions that helped to considerably improve the technical quality of the paper. This project has received funding from the EIT Digital HII project ACTIVE, and the European Union's Seventh Framework Programme for research, technological development and demonstration under grant agreement no. 607109.

\bibliographystyle{IEEEtranS}
\bibliography{main}

% Generated by IEEEtranS.bst, version: 1.12 (2007/01/11)
\begin{thebibliography}{10}
\providecommand{\url}[1]{#1}
\csname url@samestyle\endcsname
\providecommand{\newblock}{\relax}
\providecommand{\bibinfo}[2]{#2}
\providecommand{\BIBentrySTDinterwordspacing}{\spaceskip=0pt\relax}
\providecommand{\BIBentryALTinterwordstretchfactor}{4}
\providecommand{\BIBentryALTinterwordspacing}{\spaceskip=\fontdimen2\font plus
\BIBentryALTinterwordstretchfactor\fontdimen3\font minus
  \fontdimen4\font\relax}
\providecommand{\BIBforeignlanguage}[2]{{%
\expandafter\ifx\csname l@#1\endcsname\relax
\typeout{** WARNING: IEEEtranS.bst: No hyphenation pattern has been}%
\typeout{** loaded for the language `#1'. Using the pattern for}%
\typeout{** the default language instead.}%
\else
\language=\csname l@#1\endcsname
\fi
#2}}
\providecommand{\BIBdecl}{\relax}
\BIBdecl

\bibitem{Lantz:2010}
{B. Lantz, B. Heller and N. McKeown}, ``{A network in a laptop: rapid
  prototyping for software-defined networks},'' in \emph{{The Ninth ACM SIGCOMM
  Workshop on Hot Topics in Networks}}, {October} {2010}, pp. {1--6}.

\bibitem{Klein:2013}
{D. Klein and M. Jarschel}, ``{An OpenFlow extension for the OMNeT++ {INET}
  framework},'' in \emph{{6th International {ICST} Conference on Simulation
  Tools and Techniques (SimuTools '13)}}, {March} {2013}, pp. {322--329}.

\bibitem{Kreutz:2015}
{D. Kreutz, F. M. V. Ramos, P. E. Veríssimo, C. E. Rothenberg, S. Azodolmolky
  and S. Uhlig}, ``{Software-Defined Networking: A Comprehensive Survey},''
  \emph{{Proceedings of the IEEE}}, vol. {103}, no.~{1}, pp. {14--76},
  {January} {2015}.

\bibitem{OpenFlow}
\BIBentryALTinterwordspacing
{D. Pitt}, ``{Open Networking Foundation},'' {2012}. [Online]. Available:
  \url{http://opennetworking.org}
\BIBentrySTDinterwordspacing

\bibitem{Inet}
\BIBentryALTinterwordspacing
{INET Community}. {INET Framework Website}. [Online]. Available:
  \url{https://inet.omnetpp.org/}
\BIBentrySTDinterwordspacing

\bibitem{Giotis:2014}
{K. Giotis, C. Argyropoulos, G. Androulidakis, D. Kalogeras V. Maglaris},
  ``{Combining OpenFlow and sFlow for an Effective and Scalable Anomaly
  Detection and Mitigation Mechanism on SDN Environments},'' \emph{{Computer
  Networks}}, vol.~{62}, pp. {122--136}, {April} {2014}.

\bibitem{SDNsimulator}
\BIBentryALTinterwordspacing
{M. Tiloca, A. Stagkopoulou, G. Dini}, ``{INET\_SDN\_dev},'' {2016}. [Online].
  Available: \url{https://github.com/marco-tiloca-sics/INET_SDN_dev}
\BIBentrySTDinterwordspacing

\bibitem{SEA++paper}
{M. Tiloca, F. Racciatti and G. Dini}, ``{Simulative Evaluation of Security
  Attacks in Networked Critical Infrastructures},'' in \emph{{2nd International
  Workshop on Reliability and Security Aspects for Critical Infrastructure
  Protection (ReSA4CI 2015), published in Lecture Notes in Computer Science,
  LNCS 9338}}.\hskip 1em plus 0.5em minus 0.4em\relax {Springer International
  Publishing}, {September} {2015}, pp. {314--323}.

\bibitem{NS-3}
\BIBentryALTinterwordspacing
{NS-3 Project}. {NS-3 v3.16 OpenFlow switch support}. [Online]. Available:
  \url{https://www.nsnam.org/docs/release/3.16/models/html/openflow-switch.html}
\BIBentrySTDinterwordspacing

\bibitem{SDNwhitepaper}
{Open Networking Foundation}, ``{Software-Defined Networking: The New Norm for
  Networks},'' {2012}.

\bibitem{Mousavi:2015}
{S. M. Mousavi and M. St-Hilaire}, ``{Early detection of DDoS attacks against
  SDN controllers},'' in \emph{{2015 International Conference on Computing,
  Networking and Communications (ICNC 2015)}}, {February} {2015}, pp. {77--81}.

\bibitem{Oshima:2010}
{S. Oshima and T. Nakashima and T. Sueyoshi}, ``{Early DoS/DDoS Detection
  Method using Short-term Statistics},'' in \emph{{2010 International
  Conference on Complex, Intelligent and Software Intensive Systems (CISIS)}},
  {February} {2010}, pp. {168--173}.

\bibitem{Wu:2009}
{S.-Y. Wu and E. Yen}, ``{Data Mining-based Intrusion Detectors},''
  \emph{{Expert Systems with Applications}}, vol.~{36}, no.~{3}, pp.
  {5605--5612}, {April} {2009}.

\bibitem{Ahmed:2007}
{T. Ahmed, B. Oreshkin and M. Coates}, ``{Machine Learning Approaches to
  Network Anomaly Detection},'' in \emph{{Proceedings of the 2Nd USENIX
  Workshop on Tackling Computer Systems Problems with Machine Learning
  Techniques}}, ser. {SYSML'07}.\hskip 1em plus 0.5em minus 0.4em\relax
  {Berkeley, CA, USA}: {USENIX Association}, {2007}, pp. {1--6}.

\bibitem{Omnet}
\BIBentryALTinterwordspacing
A.~Vargas. {OMNeT++ Community Website}. [Online]. Available:
  \url{http://www.omnetpp.org/}
\BIBentrySTDinterwordspacing

\end{thebibliography}

\end{document}